\documentclass[preprintnumbers,10pt,nofootinbib]{revtex4}

\usepackage{amsmath,latexsym,amssymb,amsfonts}
\usepackage[pdftex]{color,graphicx}
\usepackage{bm}
\usepackage{xcolor}
\usepackage{colortbl}

\begin{document}


\title{\textbf{Isentropic process of Reissner-Nordstr{\"o}m black holes:\\ a possible excess of the entropy bound via a non-perturbative channel}}

\author{
Robert B. Mann$^{a,b,c}$\footnote{{\tt rbmann@uwaterloo.ca}}
and
Dong-han Yeom$^{a,d,e,f}$\footnote{{\tt innocent.yeom@gmail.com}}
}

\affiliation{
$^{a}$Department of Physics and Astronomy, University of Waterloo, Waterloo, ON N2L 3G1, Canada\\
$^{b}$Institute for Quantum Computing, University of Waterloo, Waterloo, ON N2L 3G1, Canada\\
$^{c}$Perimeter Institute for Theoretical Physics, Waterloo, ON N2L 2Y5, Canada\\
$^{d}$Department of Physics Education, Pusan National University, Busan 46241, Republic of Korea\\
$^{e}$Research Center for Dielectric and Advanced Matter Physics, Pusan National University, Busan 46241, Republic of Korea\\
$^{f}$Leung Center for Cosmology and Particle Astrophysics, National Taiwan University, Taipei 10617, Taiwan
}

\begin{abstract}
We study the implications of an isentropic processes applied to a Reissner-Nordstr{\"o}m black hole. This process is possible if a black hole absorbs a particle with a specific ratio of energy and charge. We show that such an absorption process is not classically allowed, not only in Einstein gravity but also in several modified gravity theories, indicating that this prohibition is quite generic. However, an isentropic absorption process is quantum mechanically allowed: the particle can penetrate the potential barrier on the event horizon. We compute the probability of this absorption process and compare it to that of semi-classical effects. Non-perturbatively,
 if this process is accumulated, it is possible that the entanglement entropy can be greater than its Bekenstein-Hawking entropy and violate the entropy-bound relation.
\end{abstract}

\maketitle

\newpage

\tableofcontents

\section{Introduction}

The information loss paradox remains one of the major unsolved research topics in modern theoretical physics \cite{Hawking:1976ra}. A black hole evaporates via Hawking radiation and disappears eventually \cite{Hawking:1975vcx}. After the black hole disappears, can we recover the information from the matter that originally formed the black hole?

If one assumes that information is carried by Hawking radiation in a semi-classical background, this will yield an inconsistency because quantum information either must be duplicated (cloned) \cite{Yeom:2009zp} or entanglement entropy relations must break down \cite{Almheiri:2012rt}. On the other hand, if information is not preserved, this is contradictory not only with holography \cite{Maldacena:1997re} but with the quantum physics that gives rise to black hole evaporation in the first place \cite{Banks:1983by}. Alternative channels for carrying away information, for example, remnants or baby universes, can be invoked \cite{Chen:2014jwq}, but these might not be generic for all situations.

A careful examination of the paradox indicates its underlying assumptions \cite{Ong:2016iwi}.
\begin{itemize}
\item[--] 1. \textit{Unitarity}: one can describe the entire life of a black hole in a unitary way, i.e., from its formation to its end state.
\item[--] 2. \textit{Local quantum field theory}: Local quantum field theory is sufficient for describing the 
Hawking radiation emitted from a black hole. 
\item[--] 3. \textit{General covariance}: A generally covariant description of gravitation is correct except in regions of spacetime where singularities are present.
\item[--] 4. \textit{Area/entropy relation}: The Bekenstein-Hawking entropy $A/4$ \cite{Bekenstein:1973ur} is proportional to the Boltzmann entropy of the black hole, where $A$ is the area of the event horizon; in a more general theory of gravity, it will be given by the Iyer-Wald formula \cite{Iyer:1994ys}.
\item[--] 5. \textit{Information observer}: There exists an observer who can measure the entanglement or information contained in Hawking radiation.
\end{itemize}
The information paradox is tantamount to the statement that these five assumptions cannot simultaneously be valid. At present, the correct answer to the information loss paradox still eludes us. However, it seems reasonable to assert the following: if the information is preserved, the corresponding black hole dynamics depends not only on semi-classical geometry but also on non-perturbative effects.

What is the meaning of non-perturbative effects? What changes would be made if we included such effects, for example, from a Euclidean path integral \cite{Chen:2022ric} or replica tricks \cite{Almheiri:2019qdq}? Here we will argue that non-perturbative effects can be expected to modify assumption 4; in other words, the Boltzmann entropy inside a black hole may be greater than its Bekenstein-Hawking entropy due to non-perturbative effects. There have been similar arguments in the literature \cite{Buoninfante:2021ijy,Bae:2020lql}, but we make a more explicit construction in this paper. Specifically, we will construct a situation in which a charged particle can be absorbed into a black hole isentropically, where the entropy of the black hole is not changed before and after the absorption, while the collapsing particle can be entangled with the outside of the black hole. This process is classically forbidden, but quantum mechanically allowed.

Our conclusion does imply an overthrow of previous well-known theorems or principles, e.g., entropy-bound relations. Semi-classical expectation values still represent the dominant contribution amongst all possible quantum effects, with non-perturbative effects typically exponentially suppressed compared to semi-classical contributions. Instead, we are pointing out that a highly suppressed process exists that violates the entropy-bound relation. Even though these processes are not dominant probabilistically, if they play some role in the information-preserving process, we need to understand them.

Our paper is organized as follows. In Sec.~\ref{sec:rad}, we discuss a radial geodesic motion in a spherically symmetric background, where we are especially interested in a charged black hole background. We further show that the classical absorption of a charged particle with the isentropic condition is impossible in Einstein gravity and quite generic modified gravity models. However, in Sec.~\ref{sec:qua}, we show that the particle can penetrate the potential barrier through a non-perturbative process. The probability might be enhanced if we think about the modified gravity effect. Finally, in Sec.~\ref{sec:con}, we discuss possible future applications.

\section{\label{sec:rad}Radial geodesic motion and the isentropic process}

\subsection{The radial geodesic equation}

We start from the general metric form of a static and spherical symmetric background:
\begin{eqnarray}
ds^{2} = - g_{00}(r) dt^{2} + g_{11}(r) dr^{2} + r^{2} d\Omega^{2},
\end{eqnarray}
where $g_{00}$ and $g_{11}$ are functions of $r$. 

Let us assume that the rest mass of a particle is $m$ and its electric charge is $q$. The asymptotic energy of the particle $E$ is
\begin{eqnarray}
E = \gamma m \sqrt{g_{00}} + q \Phi,
\end{eqnarray}
where $\gamma = 1/\sqrt{1 - v^{2}}$ is the Lorentz factor, $\sqrt{g_{00}}$ is the red-shift factor, and $q\Phi$ corresponds the electrostatic energy contribution. From this, we obtain
\begin{eqnarray}
\gamma = \frac{1}{m\sqrt{g_{00}}} \left( E - q \Phi \right)
\end{eqnarray}
and thus
\begin{eqnarray}
\gamma^{2} v^{2} = \gamma^{2} - 1 = \frac{1}{m^{2} g_{00}} \left( E - q \Phi \right)^{2} - 1.
\end{eqnarray}
The radial distance $d\ell = \sqrt{g_{11}} dr$ per proper time $d\tau$ of a particle that is moving purely radially (i.e., assuming the angular momentum equals zero) is
\begin{eqnarray}
\frac{d\ell}{d\tau} = \pm \gamma v,
\end{eqnarray}
or equivalently,
\begin{eqnarray}
\left( \frac{dr}{d\tau} \right)^{2} = \frac{\gamma^{2} v^{2}}{g_{11}} = \frac{1}{m^{2} g_{00} g_{11}} \left( E - q \Phi \right)^{2} - \frac{1}{g_{11}}\; 
\end{eqnarray}
or alternatively
\begin{eqnarray}
\left( \frac{dr}{d\tau} \right)^{2}
+ V_{\mathrm{eff}} (r) = 0,
\end{eqnarray}
where the effective potential is
\begin{eqnarray}
V_{\mathrm{eff}} (r) = \frac{\gamma^{2} v^{2}}{g_{11}} = -\frac{1}{m^{2} g_{00} g_{11}} \left( E - q \Phi \right)^{2} + \frac{1}{g_{11}}
\end{eqnarray}
valid for any theory of gravity. For metrics of the form
\begin{equation}\label{fmet}
    ds^{2} = - f(r) dt^{2} + \frac{1}{f(r)} dr^{2} + r^{2} d\Omega^{2}
\end{equation}
this becomes
\begin{eqnarray}
V_{\mathrm{eff}} (r) = - \frac{1}{m^{2}} \left( E - q\Phi \right)^{2} + f(r).
\end{eqnarray}

\subsection{Reissner-Nordstr{\"o}m black hole}

The Reissner-Nordstr{\"o}m black hole solution is
\begin{eqnarray}
g_{00}(r) = \frac{1}{g_{11}(r)} = 1 - \frac{2M}{r} + \frac{Q^{2}}{r^{2}},
\end{eqnarray}
where $M$ is the mass, and $Q$ is the electric charge of the black hole \cite{Reissner:1916cle,Nordstrom}. The event horizon $r_{+}$ is
\begin{eqnarray}
r_{+} = M + \sqrt{M^{2} - Q^{2}}.
\end{eqnarray}
The Bekenstein-Hawking entropy $S$ is
\begin{eqnarray}
S = \frac{A}{4} = \pi r_{+}^{2}.
\end{eqnarray}
The Hawking temperature $T$ and the electrostatic potential $\Phi$ are
\begin{eqnarray}
T &=& \frac{\sqrt{M^{2}-Q^{2}}}{2\pi r_{+}^{2}},\\
\Phi &=& \frac{Q}{r}.
\end{eqnarray}
From this, we obtain the effective potential of the radial geodesic motion:
\begin{eqnarray}
V_{\mathrm{eff}} (r) \equiv - \frac{1}{m^{2}} \left( E - q \frac{Q}{r} \right)^{2} + \left(1 - \frac{2M}{r} + \frac{Q^{2}}{r^{2}} \right),
\end{eqnarray}
where the classical trajectory is allowed only if $V_{\mathrm{eff}} < 0$.

\subsection{Isentropic process via particle absorption}

We can check whether an isentropic (equivalently, adiabatic) process is allowed due to the absorption of a classical (test) particle by the black hole. If this process  does not change the entropy of the black hole, then
\begin{eqnarray}
TdS = 0 = dM - \Phi dQ.
\end{eqnarray}
Substituting $dM \rightarrow E$ and $dQ \rightarrow q$, the isentropic condition becomes
\begin{eqnarray}
E - q\frac{Q}{r_{+}} = 0
\end{eqnarray}
valid for any theory of gravity coupled to electromagnetism for which
$$
\Phi = \frac{Q}{r}.
$$
With this condition, $V_{\mathrm{eff}} (r_{+}) = 0$ and
\begin{eqnarray}
\left. \frac{dV_{\mathrm{eff}}}{dr} \right|_{r_{+}} = \left. \frac{df}{dr} \right|_{r_{+}} = 4\pi T > 0.
\end{eqnarray}
Therefore, there exists $r_{1} > r_{+}$ that satisfies $V_{\mathrm{eff}} (r_{1}) = 0$. This implies that it is not possible to realize this adiabatic process via the absorption of a classical charged particle. We emphasize that this conclusion is quite generic, provided $g_{00} = g_{11}^{-1}$ and the electrostatic potential form is unchanged.

\subsection{Modified gravity example: generalized quasi-topological black holes}

As a specific example, we can consider the generalized metric function 
\begin{eqnarray}
f(r) = 1 + \frac{r^{2}}{2\alpha} \left( 1 - \sqrt{1 + \frac{8\alpha M}{r^{3}} - \frac{4\alpha Q^{2}}{r^{4}} } \right)
\end{eqnarray}
from 4-dimensional Einstein-Gauss-Bonnet (4DEGB) gravity, where $\alpha$ is a coupling constant \cite{Hennigar:2020lsl,Fernandes:2020rpa}. The horizon is located $r_{+} = M + \sqrt{M^{2} - Q^{2} -\alpha}$, and the solution will be restored to the Reissner-Nordstr{\"o}m solution if $\alpha$ goes to zero.

\begin{figure}[h]
\centering
\includegraphics[scale=0.6]{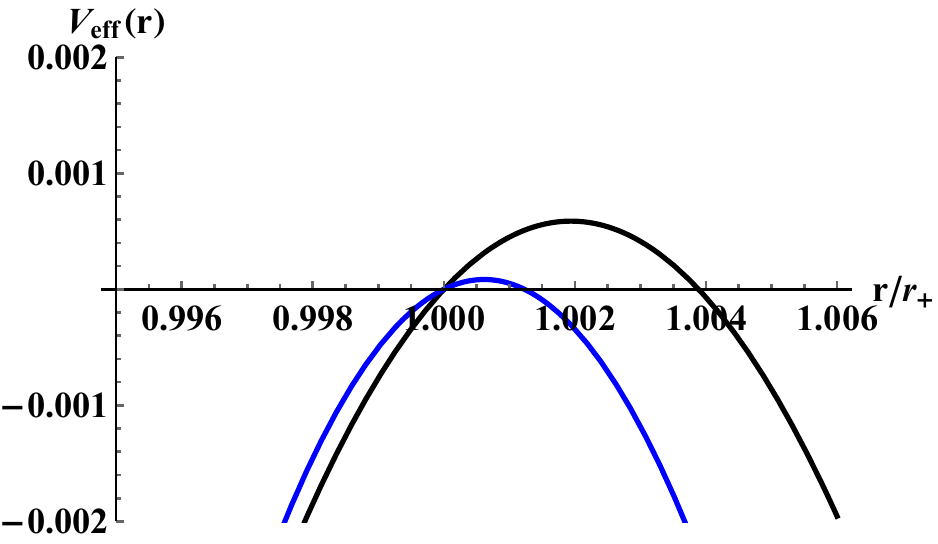}
\caption{$V_{\mathrm{eff}}(r)$ as a function of $r/r_{+}$, where the black curve is for Einstein gravity, and the blue curve is for modified gravity, where $M = 1$, $Q=0.9$, $m=0.0002$, $q=0.002$, and $\alpha = 0.15$.}
\label{fig:exm}
\end{figure}

\begin{figure}[h]
\centering
\includegraphics[scale=0.3]{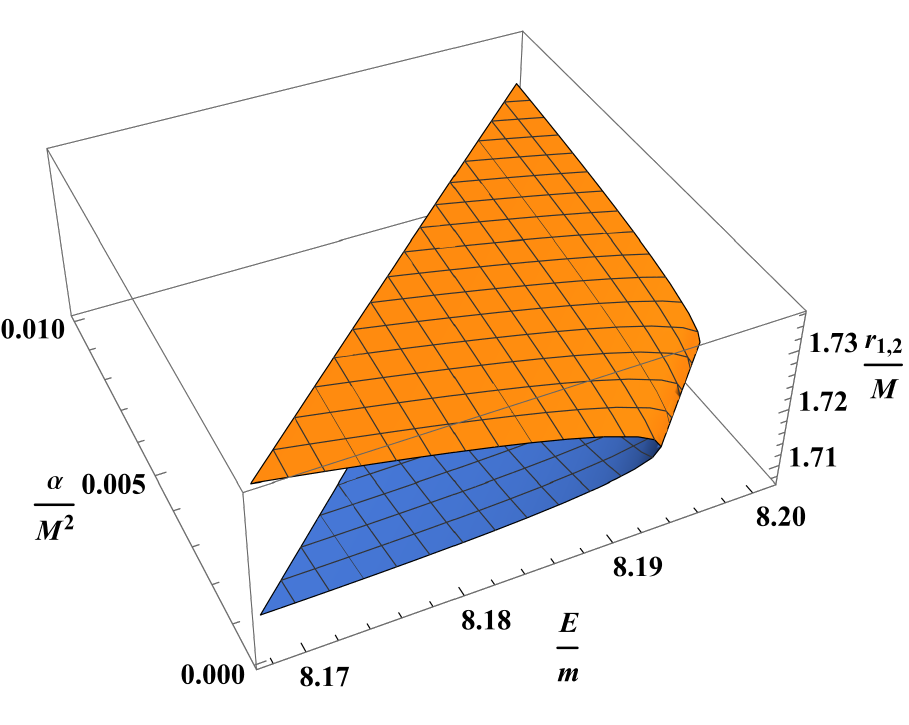}
\includegraphics[scale=0.3]{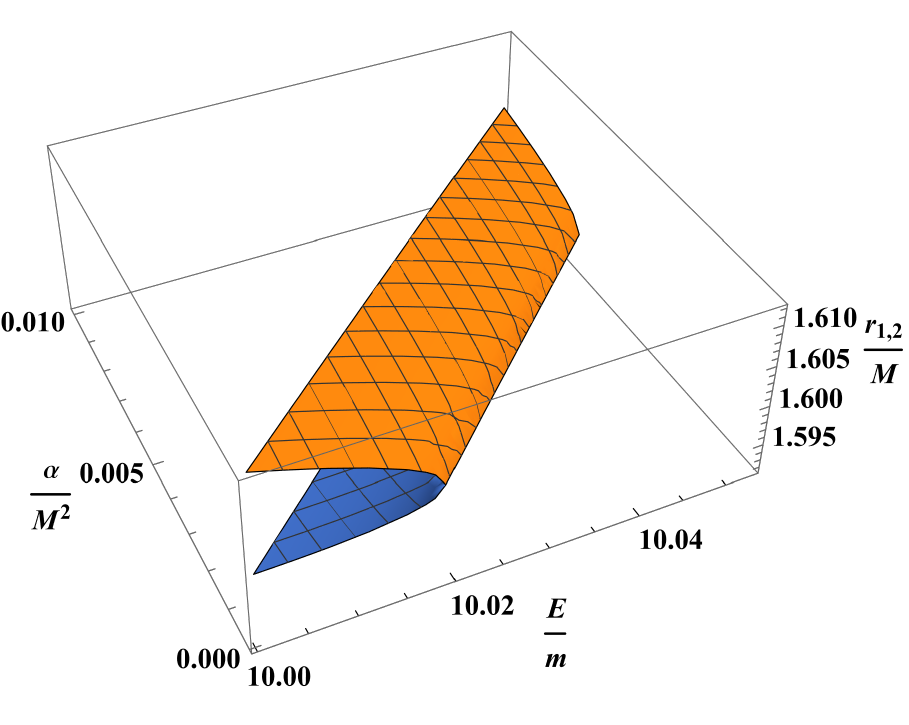}
\includegraphics[scale=0.3]{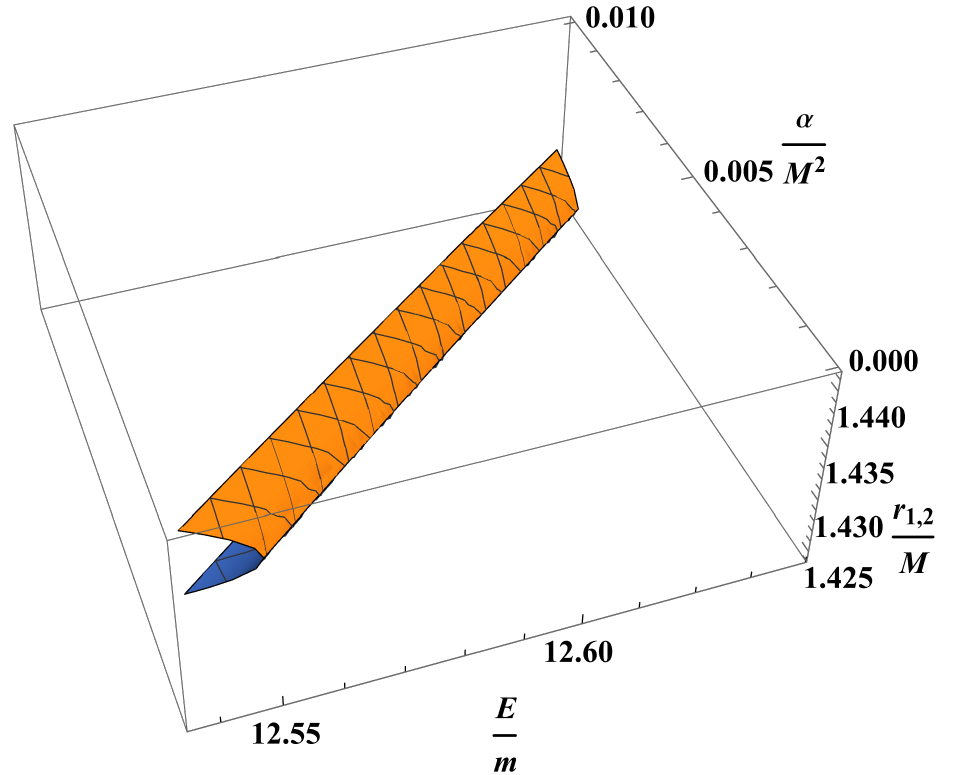}
\caption{$r_{1,2}/M$ satisfying $V_{\mathrm{eff}}(r_{1,2}) = 0$ and $r_{1} > r_{2}$, by varying $E/m$, $\alpha^{2}/M$, and fixing $q/m = 20$, $Q/M = 0.7$ (left), $0.8$ (middle), and $0.9$ (right). The left cutting edge satisfies the isentropic condition.}
\label{fig:params}
\end{figure}

However, there is a significant difference compared to Einstein gravity. Fig.~\ref{fig:exm} compares the effective potential of the Einstein and modified gravity models. For fixed $Q/M$ and $q/m$, we see from Fig.~\ref{fig:exm} that increasing $\alpha$ decreases the height of the potential barrier. Fig.~\ref{fig:params} shows that the potential barrier has not disappeared; removing it entails violating the isentropic condition.

\section{\label{sec:qua}Quantum mechanical absorption}

In the adiabatic case, we have shown that classical absorption is impossible. However, quantum tunneling could allow the particle to penetrate the potential barrier. If so, the isentropic process is possible because the black hole can absorb particles consistent with the isentropic condition if we repeatedly send particles.

\subsection{WKB approximation}

The barrier penetration probability can be computed by using the WKB approximation.
For a metric of the form \eqref{fmet}, we set  $d\theta = d\phi = 0$ for a  radial geodesic, obtaining
\begin{eqnarray}
L = -f \dot{t}^{2} + \frac{\dot{r}^{2}}{f} = -1,
\end{eqnarray}
where $L$ is the Lagrangian of a timelike geodesic.

Let us define $r_{1} < r_{2}$ satisfying $V_{\mathrm{eff}} (r_{1,2}) = 0$. For $r_{1} < r < r_{2}$,  a particle cannot be located classically. The tunneling rate, according to the WKB approximation, is
\begin{eqnarray}
\Gamma \simeq e^{-2 S_{\mathrm{E}}},
\end{eqnarray}
where   the Euclidean action that connects between $r_{1}$ and $r_{2}$ is 
\begin{eqnarray}
S_{\mathrm{E}} &=& \int L_{\mathrm{E}} d\tau \\
&=& \int_{r_{1}}^{r_{2}} \frac{1}{(dr/d\tau)} dr\\
&=& \int_{r_{1}}^{r_{2}} \frac{1}{\sqrt{V_{\mathrm{eff}}(r)}} dr.
\end{eqnarray}

\begin{figure}
\centering
\includegraphics[scale=0.6]{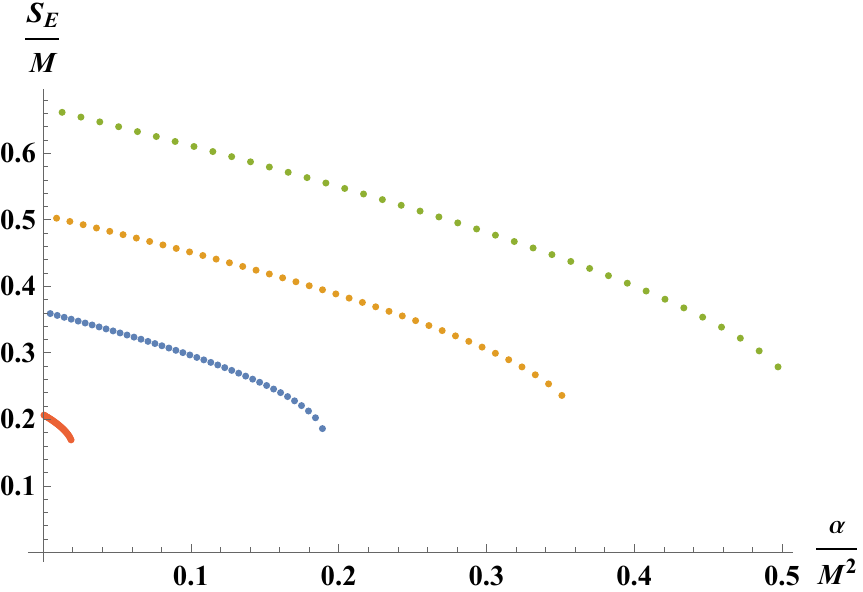}
\caption{$S_{\mathrm{E}}/M$ of isentropic conditions by varying $\alpha^{2}/M$, and fixing $q/m = 20$ and $Q/M = 0.99$ (red), $0.9$ (blue), $0.8$ (orange), $0.7$ (green), respectively.}
\label{fig:2}
\end{figure}

Fig.~\ref{fig:2} shows that as $\alpha$ increases, $S_E$ decreases, so the probability of tunneling through the barrier increases, commensurate with the height of the potential barrier decreasing. Since $\alpha$ has a limiting value $M^{2} - Q^{2}$, the Euclidean action cannot decrease indefinitely. However, the tunneling probability is approximately
\begin{eqnarray}
\Gamma \simeq e^{-\beta M},
\end{eqnarray}
where $\beta$ is an order one constant that may depend on $\alpha$. This is comparable to the Boltzmann factor of Hawking radiation:
\begin{eqnarray}
\Gamma \simeq e^{-\frac{\omega}{T}} \simeq e^{-m M},
\end{eqnarray}
where $\omega$ is the energy of the emitted particle (approximately, $\sim m$) and $T \sim 1/M$ is the Hawking temperature. Note that $m \ll 1$ in   Planck units, and hence, the tunneling probability of such an adiabatic process is usually subdominant compared to Hawking radiation or semi-classical effects. However, it is fair to say that such a suppression is not too small to be totally neglected.

\subsection{Thermodynamic stability}

Let us evaluate the free energy functions:
\begin{eqnarray}
F &\equiv& M - ST,\\
G &\equiv& M - ST - Q\Phi.
\end{eqnarray}
Note that $S$ and $r_{+}$ are constant in isentropic processes. In Einstein gravity, one can evaluate
\begin{eqnarray}
F &=& M - \frac{1}{2} \sqrt{M^{2} - Q^{2}} \\
  &=& - \frac{r_{+}}{2} + \frac{3}{2} M,\\
G &=& - \frac{r_{+}}{2} + \frac{3}{2} M - \frac{Q^{2}}{r_{+}} \\
&=& - \frac{r_{+}}{2} + \frac{3}{2} M + \frac{(r_{+} - M)^{2} - M^{2}}{r_{+}} \\
  &=& + \frac{r_{+}}{2} - \frac{1}{2} M.
\end{eqnarray}
Therefore, if particle absorption is isentropic, then $dM > 0$, and hence, $F$ increases and $G$ decreases.

For 4DEGB gravity, we find 
\begin{eqnarray}
Q = \sqrt{M^{2} - \alpha - (r_{+} - M)^{2}},
\end{eqnarray}
since $r_{+}$ is  constant; one can simplify
\begin{eqnarray}
F &=& - \frac{r_{+}}{2 + 4 \alpha/r_{+}^{2}} + \left( \frac{3 + 4\alpha / r_{+}^{2}}{2 + 4 \alpha/r_{+}^{2}} \right) M,\\
G &=& + \frac{r_{+} (1 + 6\alpha / r_{+}^{2} + 4 \alpha^{2}/ r_{+}^{4})}{2 + 4\alpha/r_{+}^{2}} - \frac{M}{2},
\end{eqnarray}
and so the qualitative behaviours are similar to those in Einstein gravity.

\subsection{Can the entropy-bound relation be violated?}

If we accept non-perturbative processes, adiabatic absorption is possible. When the black hole is maximally entangled with its background, i.e., when the Bekenstein-Hawking entropy is the same as its entanglement entropy, let us assume that we send a particle entangled outside. Because of the isentropic condition, its areal entropy is the same. However, the entanglement entropy between inside and outside the horizon must increase.

The entanglement entropy is always smaller than its Boltzmann entropy. Therefore, it is inevitable to accept the possibility that the Boltzmann entropy can be greater than the Bekenstein-Hawking entropy if one allows non-perturbative effects.

\section{\label{sec:con}Conclusion}

In this paper, we investigated the possibility that a particle can be absorbed by a charged black hole while keeping its entropy constant. We have shown that such an isentropic process is prohibited classically but allowed quantum mechanically. The tunneling probability is exponentially suppressed compared to the dominant semi-classical effects, e.g., Hawking radiation; however, it is fair to say that the tunneling probability is non-zero and even slightly enhanced if we consider modified gravity effects.

We thus conclude that \textit{a non-perturbative quantum process can violate the entropy-bound relation}. The non-perturbative process is suppressed and ignored by usual semi-classical computations that include perturbative or thermal quantum effects. Consequently, previously proven entropy-bound relations are still consistent \cite{Bekenstein:1980jp,Bousso:2002ju}, and one can reasonably ignore such non-perturbative effects if we average over all possible contributions. However, if we believe that non-perturbative effects are essential for solving the information loss paradox \cite{Maldacena:2001kr,Hawking:2005kf,Sasaki:2014spa}, we should open the possibility that they can provide crucial insights \cite{Chen:2015gux,Chen:2016nvj}.

What happens if isentropic absorption accumulates? Such an object will have more Boltzmann entropy than its Bekenstein-Hawking entropy, which is sometimes called a \textit{monster}. How can information be rescued from such a semi-classical monster \cite{Chen:2014jwq}? In this case, should there be island contributions automatically \cite{Almheiri:2019hni}? Will the island contributions destroy a semi-classical background or not \cite{Marolf:2020rpm}? If we accept the possibility of the existence of such an object, such questions inevitably present themselves. These are unresolved and interesting questions, and we leave them for future work.

\section*{Acknowledgements}
The authors would like to thank Don Page for discussing this work at an early stage. DY was supported by the National Research Foundation of Korea (Grant Nos. : 2021R1C1C1008622, 2021R1A4A5031460).


\begin{thebibliography}{200}

\bibitem{Hawking:1976ra}
S.~W.~Hawking,
Phys. Rev. D \textbf{14}, 2460-2473 (1976).

\bibitem{Hawking:1975vcx}
S.~W.~Hawking,
Commun. Math. Phys. \textbf{43}, 199-220 (1975)
[erratum: Commun. Math. Phys. \textbf{46}, 206 (1976)].

\bibitem{Yeom:2009zp}
D.~Yeom and H.~Zoe,
Int. J. Mod. Phys. A \textbf{26}, 3287-3314 (2011)
[arXiv:0907.0677 [hep-th]].

\bibitem{Almheiri:2012rt}
A.~Almheiri, D.~Marolf, J.~Polchinski and J.~Sully,
JHEP \textbf{02}, 062 (2013)
[arXiv:1207.3123 [hep-th]].

\bibitem{Maldacena:1997re}
J.~M.~Maldacena,
Adv. Theor. Math. Phys. \textbf{2}, 231-252 (1998)
[arXiv:hep-th/9711200 [hep-th]].

\bibitem{Banks:1983by}
T.~Banks, L.~Susskind and M.~E.~Peskin,
Nucl. Phys. B \textbf{244}, 125-134 (1984).

\bibitem{Chen:2014jwq}
P.~Chen, Y.~C.~Ong and D.~Yeom,
Phys. Rept. \textbf{603}, 1-45 (2015)
[arXiv:1412.8366 [gr-qc]].

\bibitem{Ong:2016iwi}
Y.~C.~Ong and D.~Yeom,
``Summary of Parallel Session: Black Hole Evaporation and Information Loss Paradox,''
Proceedings of the 2nd LeCosPA Symposium
[arXiv:1602.06600 [hep-th]].

\bibitem{Bekenstein:1973ur}
J.~D.~Bekenstein,
Phys. Rev. D \textbf{7}, 2333-2346 (1973).

\bibitem{Iyer:1994ys}
V.~Iyer and R.~M.~Wald,
Phys. Rev. D \textbf{50}, 846-864 (1994)
[arXiv:gr-qc/9403028 [gr-qc]].

\bibitem{Chen:2022ric}
P.~Chen, M.~Sasaki, D.~Yeom and J.~Yoon,
Entropy \textbf{25}, no.12, 1663 (2023)
[arXiv:2206.10251 [gr-qc]].

\bibitem{Almheiri:2019qdq}
A.~Almheiri, T.~Hartman, J.~Maldacena, E.~Shaghoulian and A.~Tajdini,
JHEP \textbf{05}, 013 (2020)
[arXiv:1911.12333 [hep-th]].

\bibitem{Buoninfante:2021ijy}
L.~Buoninfante, F.~Di Filippo and S.~Mukohyama,
JHEP \textbf{10}, 081 (2021)
[arXiv:2107.05662 [hep-th]].

\bibitem{Bae:2020lql}
J.~M.~Bae, D.~J.~Lee, D.~Yeom and H.~Zoe,
Symmetry \textbf{14}, no.8, 1649 (2022)
[arXiv:2002.03543 [hep-th]].

\bibitem{Reissner:1916cle}
H.~Reissner,
Annalen Phys. \textbf{355}, no.9, 106-120 (1916).

\bibitem{Nordstrom}
G.~Nordström, Verhandl. Koninkl. Ned. Akad. Wetenschap., Afdel. Natuurk. \textbf{26}, 1201–1208 (1918).

\bibitem{Hennigar:2020lsl}
R.~A.~Hennigar, D.~Kubiz\v{n}\'ak, R.~B.~Mann and C.~Pollack,
JHEP \textbf{07}, 027 (2020)
[arXiv:2004.09472 [gr-qc]].

\bibitem{Fernandes:2020rpa}
P.~G.~S.~Fernandes,
Phys. Lett. B \textbf{805}, 135468 (2020)
[arXiv:2003.05491 [gr-qc]].

\bibitem{Bekenstein:1980jp}
J.~D.~Bekenstein,
Phys. Rev. D \textbf{23}, 287 (1981).

\bibitem{Bousso:2002ju}
R.~Bousso,
Rev. Mod. Phys. \textbf{74}, 825-874 (2002)
[arXiv:hep-th/0203101 [hep-th]].

\bibitem{Maldacena:2001kr}
J.~M.~Maldacena,
JHEP \textbf{04}, 021 (2003)
[arXiv:hep-th/0106112 [hep-th]].

\bibitem{Hawking:2005kf}
S.~W.~Hawking,
Phys. Rev. D \textbf{72}, 084013 (2005)
[arXiv:hep-th/0507171 [hep-th]].

\bibitem{Sasaki:2014spa}
M.~Sasaki and D.~Yeom,
JHEP \textbf{12}, 155 (2014)
[arXiv:1404.1565 [hep-th]].

\bibitem{Chen:2015gux}
P.~Chen, Y.~C.~Ong, D.~N.~Page, M.~Sasaki and D.~Yeom,
Phys. Rev. Lett. \textbf{116}, no.16, 161304 (2016)
[arXiv:1511.05695 [hep-th]].

\bibitem{Chen:2016nvj}
P.~Chen, C.~H.~Wu and D.~Yeom,
JCAP \textbf{06}, 040 (2017)
[arXiv:1608.08695 [hep-th]].

\bibitem{Almheiri:2019hni}
A.~Almheiri, R.~Mahajan, J.~Maldacena and Y.~Zhao,
JHEP \textbf{03}, 149 (2020)
[arXiv:1908.10996 [hep-th]].

\bibitem{Marolf:2020rpm}
D.~Marolf and H.~Maxfield,
JHEP \textbf{04}, 272 (2021)
[arXiv:2010.06602 [hep-th]].






\end{thebibliography}
\end{document}